\documentclass[eqsecnum,pra,twocolumn,aps]{revtex4}
\usepackage{graphicx}
\usepackage{graphicx}
\usepackage{epsfig}
\usepackage{dcolumn}
\usepackage{bm}
\usepackage{bbm}
\usepackage{amsmath}
\usepackage{graphicx}
\usepackage{amsfonts}
\usepackage{hyperref}
\usepackage{amssymb}
\usepackage{color}
\usepackage[up]{subfigure}
\newtheorem{thm}{Theorem}

\newcommand{\be}{\begin{equation}}
\newcommand{\ee}{\end{equation}}
\newcommand{\bc}{\begin{center}}
\newcommand{\ec}{\end{center}}
\newcommand{\bea}{\begin{eqnarray}}
\newcommand{\eea}{\end{eqnarray}}

\begin{document}
\title{Quantumness of noisy quantum walks: a comparison between 
measurement-induced disturbance and quantum discord}
\author{Balaji R. Rao}
\author{R. Srikanth}
\email{srik@poornaprajna.org}
\affiliation{Poornaprajna Institute of Scientific Research,
Sadashivnagar, Bengaluru- 560 080.}
\author{C. M. Chandrashekar}
\email{chandru@imsc.res.in}
\affiliation{Center for Quantum Sciences, The Institute of Mathematical Sciences, Chennai 600113, India.}
\author{Subhashish Banerjee}
\affiliation{Indian Institute of Technology Rajasthan, Jodhpur 342011, India.}


\begin{abstract}
Noisy quantum walks are studied  from the perspective of comparing 
their    quantumness   as   defined    by   two    popular   measures,
measurement-induced disturbance (MID)  and quantum discord (QD). While
the former has  an operational definition, unlike the  latter, it also
tends to overestimate non-classicality because of a lack of optimization
over local measurements.  Applied to quantum walks, we
find that  MID, while acting as a loose upper bound on
QD, still tends to reflect correctly the trends in the  behavior of 
the latter. However, there are regimes where its behavior is not
indicative of non-classicality: in particular, we find an instance
where MID increases with the application of noise, 
where we expect a reduction of quantumness.
\end{abstract}

\maketitle

\section{Introduction}


Many useful quantities in  quantum information theory (such as various
quantifications  of entanglement and channel capacities) 
lack  an operational  definition. Quantifying the degree of nonclassicality
or quantumness in a state is one such. Intuitively, we expect that
entanglement captures all of the nonclassicality in a correlation.
We now know that in general this is not the case 
and that, for mixed states, non-classicality,
nonlocality, and entanglement are not identical. 
 
In the  case of quantifying  the quantumness of a bipartite state,  
following the  proposal of
quantum discord (QD) \cite{OZ01}, which requires an extremization over
local  measurement strategies,  measurement-induced  disturbance (MID)
\cite{L08}  was  proposed as  an  operational  measure. Recently,
QD has received several operational interpretations, 
in terms of the efficiency of Maxwell's demon \cite{Zur03+},
the entanglement consumed \cite{winter} or quantum communication 
\cite{madhok} during
state merging, and distillable entanglement in quantum measurement
\cite{strel}. However, the difficult posed by the required
optimization remains. In contrast, MID
requires no such optimization, instead it uses the local measurement
strategy defined by the diagonalization of the 
reduced density operators. 
If MID were a good indicator of non-classicality, in particular, if it correctly 
reflected the  behavior of  QD, we would have a happy instance of
an operational proxy for genuine non-classicality. However,
Ref. \cite{GPA10} has reported several difficulties with the use 
of MID for a two-qubit system. In particular, there are states of
vanishing (symmetrized) discord for which MID is maximal.
One way to ameliorate the performance of MID is to optimize it over all
possible local measurements \cite{GPA10}.
In this work, we compare these two indicators of non-classicality  
by applying them to unitary  and  noisy  discrete-time quantum  walk
(DTQW), treated as a $(2 \times k)$-dimensional system.
 
Quantum walks (QWs) \cite{ADZ93,  DM96}, which are the quantum analogs
of  classical random  walks (CRWs),  have been  extensively  studied as a 
quantum  algorithm  \cite{FG98,  ABN01,  NV01,  Amb03,  CCD03,  SKB03,
  AKR05}, to  demonstrate coherent control over  atoms \cite{CL08}, to
explain phenomena  such as the breakdown of an electric-field  driven system
\cite{OKA05},  and as direct experimental  evidence  for wavelike  energy
transfer      within      photosynthetic     systems      \cite{ECR07,
  MRL08}. Decoherence in a QW and the transition of a QW  to a CRW is quite
important from the viewpoint of practical implementation, and it has been
studied  by various  authors \cite{KT03,  BCA03, Ken06,  CSB07, BSC08,
  SBC10, LP10}.  In particular,  in Refs.  \cite{CSB07, BSC08, SBC10},
we   investigated   some  qualitatively   different   ways  in   which
environmental effects  suppress quantum superposition in a QW  on a line
and on an $n-$cycle. 

This report is  arranged as follows. In Sec.   \ref{dtqw} we 
briefly introduce the DTQW model  on a line and on an $n-$cycle as
well as  the noise channel used
for  our  study.   In  Sec.   \ref{ququ} we  compare and contrast
the quantumness
of a QW subjected to noise, as computed by QD and MID,
to quantify  the quantumness  and investigate  the proximity  of
the outcomes using the two  measures,  MID  and QD.  Finally,  we conclude  in
Sec. \ref{conc}.

\section{Discrete-time quantum walk on a line and an $n-$cycle}
\label{dtqw}

A DTQW  in  one dimension is  modeled  as  a  particle consisting  of  a
two-level coin  (a qubit)  existing in the  Hilbert space  ${\cal H}_c$,
spanned  by $|0\rangle$  and  $|1\rangle$, and  a  position degree  of
freedom  existing  in  the   Hilbert  space  ${\cal  H}_p$,  spanned  by
$|\psi_x\rangle$, where $x \in {\mathbbm  I}$, the set of integers. In
an  $n$-cycle walk,  there  are  only $n$  allowed  positions, and  in
addition  the   periodic  boundary  condition  $|\psi_x\rangle=|\psi_{x
\;{\rm mod}\;n}\rangle$ is imposed.  For our study, a  $t$-step coined QW is generated
by  iteratively applying  a unitary  operation $W$,  which acts  on the
Hilbert     space    ${\cal     H}_c\otimes    {\cal     H}_p$: 
 \be
|\Psi_t\rangle=W^t|\Psi_{in}\rangle,  \ee  where $|\Psi_{in}\rangle  = \frac{1}{\sqrt 2}
(|0\rangle+   i |1\rangle)\otimes |\psi_0\rangle$  is  an initial  state  of  the
particle and $W\equiv U(B \otimes  {\mathbbm 1})$, where $U(2) \ni B =
B_{\theta}       \equiv      \left(      \begin{array}{clcr}
  \cos(\theta)      &     &    $~~~$ \sin(\theta)
  \\ \sin(\theta) & & - \cos(\theta)
\end{array} \right)$ is  the  coin operation.   $U$ is  the controlled-shift
operation         
   \be 
        U\equiv           |0\rangle\langle
0|\otimes\sum_x|\psi_x-1\rangle\langle   \psi_x|   +  |1\rangle\langle
1|\otimes\sum_x|\psi_x+1\rangle\langle \psi_x|.   
\ee
For an $n-$cycle, $|\psi_x-1\rangle$ and $|\psi_x+1\rangle$ are replaced by $|\psi_{x-1~{\rm  mod}~n}\rangle$ and $|\psi_{x+1~{\rm   mod}~n}\rangle$, respectively. The probability  of finding the
particle at  site $x$ after  $t$ steps is  given by $p(x,t)  = \langle
\psi_x|{\rm Tr}_c (|\Psi_t\rangle\langle\Psi_t|)|\psi_x\rangle$.

To quantify quantumness when noise is applied to a DTQW, we will consider the amplitude-damping channel \cite{NC00} parametrized by $\lambda$ which has the following operator-sum representation :
\bea
\label{noisechan}
E_0 \equiv
\begin{bmatrix}
\sqrt{1 - \lambda} & 0 \\
0 & 1
\end{bmatrix}
~~~;~~~~
E_1 \equiv
\begin{bmatrix}
0 & 0 \\
0 & \sqrt{\lambda}
\end{bmatrix}
\eea
where $\lambda$ ranges from the noiseless case (0) to that of maximum noise (1). 
More general noise models can be used, such as a dissipative interaction in the presence of a squeezed thermal bath
\cite{SB08}, but the above simple model captures all the essential physics, and is hence found to be sufficient for
present purposes.

\begin{figure}
\includegraphics[width=8.9cm]{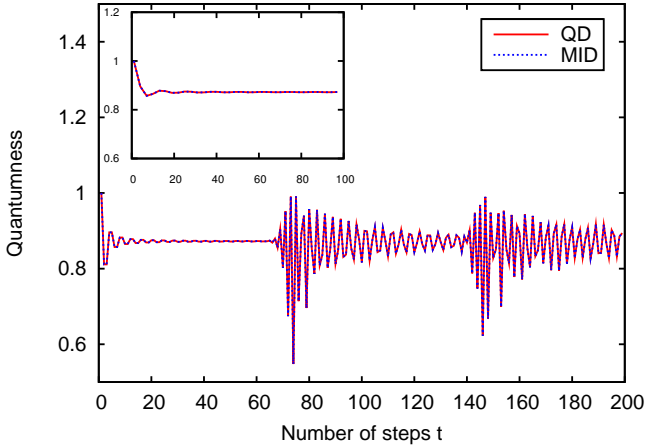}
\caption{(Color online) QD and MID for a unitary walk using $B_{\pi/2}$ as a quantum coin operation on a 51-cycle (inset is for a 100-iteration walk on a line).  For a noiseless walk, the quantumness using MID and QD is the same (see Theorem \ref{thm:Q}).}
\label{fig:1}
\end{figure}

\begin{figure}
\includegraphics[width=8.9cm]{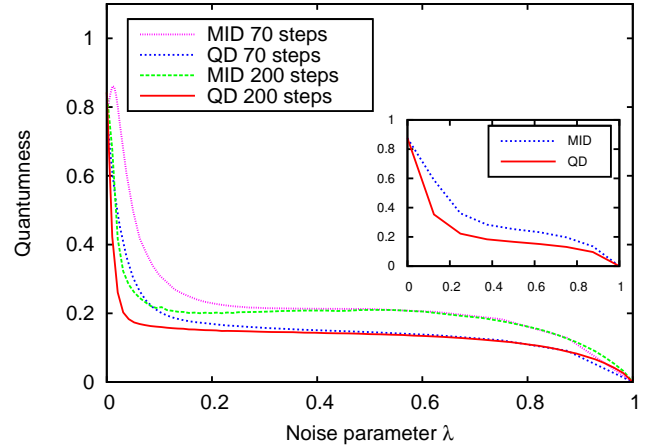}
\caption{(Color online) QD and MID for a quantum walk with the increase of noise level due to an amplitude-damping channel on a 51-cycle,
 after 70- and 200-iterations, respectively (inset is for a walk on a line after 100-iterations).  Owing to the closed, periodic dynamics 
in the $n-$cycle, the effect of very little noise is amplified leading to a steep reduction in the quantumness. 
We note that MID and QD follow a similar trend. In general, ${\rm QD} \leq {\rm MID}$ \cite{GPA10}.}
\label{fig:2}
\end{figure}

\section{Quantfying quantumness}
\label{ququ}

A number of measures for quantifying quantumness exist
\cite{OZ01, Ved03, L08, MPSVW10, GBB10}, most of
which are not operationally defined. Except in the simplest cases,
extensive numerics would be needed. From these, we selected
measurement-induced disturbance (MID) \cite{L08}, which has an
operational definition, and quantum discord (QD) \cite{OZ01},
which involves extremization over measurement strategies. 
We consider  the  classicalization of a QW  on a  line and on an
$n-$cycle under  the influence  of the amplitude-damping channel
Eq. (\ref{noisechan}).

\paragraph{Measurement-induced disturbance.}

Given a bipartite state $\rho$ existing in the 
Hilbert space ${\cal H}_C \otimes {\cal H}_P$, let the reduced density 
matrices be denoted by $\rho_C$  and $\rho_P$.
Let $\rho_C  = \sum_i
p_C^i\Pi_C^i$ and $\rho_P = \sum_j p_P^j \Pi_P^j$. The measurement
induced by the spectral resolution of the reduced states is
\begin{equation}
\label{eq:Pi}
\Pi(\rho) \equiv \sum_{j,k} \Pi_C^j \otimes \Pi_P^k \rho
\Pi_C^j \otimes \Pi_P^k,
\end{equation}
which may be  considered classical in the sense that there
is a  (unique) local measurement  strategy, namely, $\Pi$,  that leaves
$\Pi(\rho)$ unchanged. This strategy is  special in that it produces a
classical state in $\rho$ while keeping the reduced states invariant.

According to Luo  \cite{L08}, a reasonable measure of quantumness is
MID, given by
\begin{equation}
\label{eq:Q}
Q(\rho) = I(\rho) - I[\Pi(\rho)],
\end{equation}
where $I(\cdot)$ is mutual information. Accordingly, Eq. (\ref{eq:Q})
is interpreted as the difference between the total and classical correlations.

\paragraph{Quantum discord.}

Quantum discord \cite{OZ01} is given by:
\begin{eqnarray}
{\cal D}(P|C) &=& {\cal I }(P:C)^Q - {\cal J(P:C)}_{\{\Pi_j^C\}}^Q\\
&=& S(C) - S(P, C) + S(P | \{N_j^C\})\\
&=& S(P | \{N_j^C\}) - S (P | C) ,\label{eq:disc_scond}
\end{eqnarray}
where $ S(P | \{N_j^C\}) = \sum_{j}{p_j S(\rho_{X | N_j^C})}$.
$\rho_{X | N_j^C} = {\rm Tr}_C[{\mathbb I}_P \otimes N_j^C \rho_{P, C}] /$ Tr$[N_j^C \rho_{P, C}]$
is the state of $P$ after outcome $N_j^C$.
This is in general computationally very intensive. However, 
it has been shown that for qubit systems it suffices to consider
rank-1 positive operator valued measures (POVMs) \cite{Dat08}, which for qubits reduces
 to projective measurements.

We have numerically evaluated ${\cal D}(P|C)$ by minimizing Eq. \ref{eq:disc_scond}
by performing projective measurement over all bases for C parametrized by
$\alpha$ and $\beta$ : $\{\cos(\alpha)|0\rangle + e^{i\beta}\sin(\alpha)|1\rangle,
e^{-i\beta}\sin(\alpha)|0\rangle - \cos(\alpha)|1\rangle\}$.
Because of Theorem \ref{thm:Q} below, a comparison of QD and MID is
interesting only for mixed states.

\begin{thm}
\label{thm:Q}
For pure states, MID, QD and entanglement are identical.
\end{thm}

\begin{eqnarray*}
{\cal D}(P|C) &=& S(P | \{N_j^C\}) - S (P | C) \\
&=& S(P | \{N_j^C\}) - S (P,C) + S(C)\\
&=& S(P | \{N_j^C\}) + S(C).
\end{eqnarray*}

{\bf Proof.} The expression $P | \{N_j^C\}$ is the state of $P$ after $C$ is measured. In
the case of entangled pure bipartite states, 
by virtue of Schmidt decomposition, 
when the outcome of measuring C is known,
the state of P after measuring C is also exactly known and 
hence is pure. Therefore $S(P | \{N_j^C\})$ = 0. 
Hence, the expression for ${\cal D}(P|C)$ 
reduces to $S(P)=S(C)$ 
in the pure case.
Again, by Eq. (\ref{eq:Q}), MID equals $2S(C) - S(C)= S(C)$, as does
entanglement \cite{L08}. 
\hfill $\blacksquare$

\begin{figure}
\includegraphics[width=8.9cm]{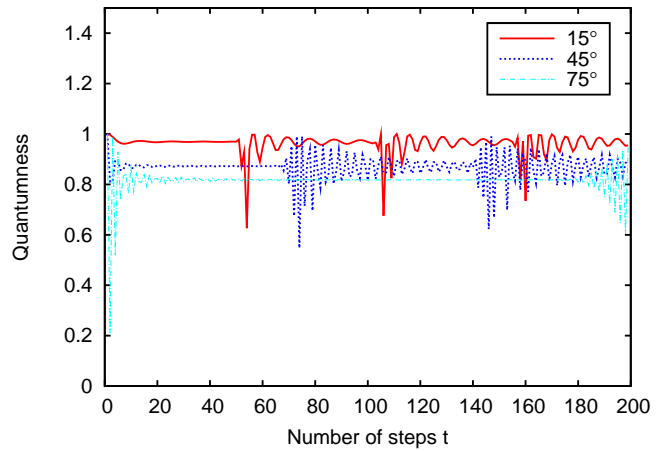}
\caption{(Color online) The quantumness using MID and QD, for a unitary walk on a 51-cycle with different coin parameters $\theta$ in $B_{\theta}$, is the same. We note 
that the frequency of dominant oscillations falls with $\theta$, a behavior we expect from the fact that
the speed of a wave packet is proportional to $\sqrt{\cos\theta}$ \cite{CBS10}.}
\label{fig:3}
\end{figure}
\begin{figure}
\includegraphics[width=8.9cm]{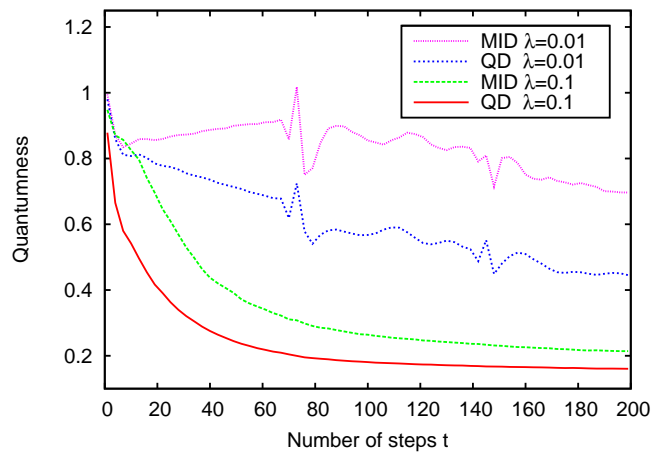}
\caption{(Color online) Quantumness on a walk with an increase in noise level on a 51-cycle using $B_{\pi/4}$ as the coin operation. 
We note that, although both quantumness measures show similar trends including with  fluctuations, in the regime of $t \sim 10$ to $t\sim 60$, MID increases (for $\lambda = 0.01$), whereas QD evinces the expected behavior.}
\label{fig:4}
\end{figure}

Two simple consequences are that for pure bipartite systems, entanglement 
captures all of the quantumness, and that QD is symmetric in this case. 
For mixed states, the situation is of course complicated. 
One fact, however, is the following result.
\begin{thm}
QD $\leq$ MID.
\label{thm:qdltmid}
\end{thm}
{\bf Proof.} 
Noting that $\rho_A = {\rm Tr}_B(\rho_{AB}) = {\rm Tr}_B(\Pi(\rho_{AB}))$,
we find $Q(\rho_{CP}) = S_\Pi(P|C) - S(P|C)$, where $S_\Pi(P|C)$ is the
conditional entropy evaluated on $\Pi(\rho)$, in view of Eq. (\ref{eq:Q}). 
Comparing this with (\ref{eq:disc_scond}) 
we find that $Q(\rho_{CP})-{\cal D}(P|C) = S_\Pi(P|C) -
S(P | \{N_j^C\})$, which is always positive for the following
reason. Clearly, $S_\Pi(P | \{N_j^C\}) \ge S(P | \{N_j^C\})$.
Now, $S_\Pi(P|C) = S(\sum_{j,k}p(j,k)|j,k\rangle_{PC} {_{PC}}\langle j,k|)
- S(\sum_jp(j)|j\rangle_C {_C}\langle j|) = 
-\sum_{j,k} p(k) p(j|k) \ln(p(j|k)) = \sum_k p(k)S_\Pi(P|C)_k
= S_\Pi(P| \{ E_j\})$, 
where $p(j,k)$ is the joint probability of outcomes $j$ and $k$ by 
measuring $\Pi(\rho_{PC})$ in the eigenbases $E_j$ of their respective reduced 
density operators, $p(j) \equiv \sum_k p(j,k)$,
$p(j|k) = p(j,k)/p(k)$
and $S_\Pi(P| \{ E_j\})$ is the average uncertainty in the first register by measuring the second register
in the diagonal basis of the latter's density operator. Clearly,   
$S_\Pi(P| \{ E_j\}) \ge S_\Pi(P | \{N_j^C\})$, and we have the required result.
\hfill $\blacksquare$

Figure \ref{fig:1} depicts QD and MID for a unitary walk for pure states, which are identical in this case
as noted in Theorem \ref{thm:Q}. We note that whereas the quantumness for a walk on a line stabilizes eventually,
that for a walk on a cycle shows a periodic increase in quantumness, 
which is associated with `crossovers',
where the left- and right-moving partial waves interfere. 
Figure \ref{fig:2} shows the expected decrease of
quantumness with noise, for both linear and cyclic walks. While
MID is seen to upper-bound QD everywhere (except at extremal points, where they
are identical), it still tends to reproduce the features of the latter's plot,
such as the steep fall and the plateau thereafter.

Figure \ref{fig:3} depicts the $\theta$-dependence of the periodicity of the 
crossovers of the left- and right- moving components of the walk. This may be
understood in terms of the wave-packet dynamics implied by the walk. In Ref. \cite{CBS10}, it was shown
that the wave velocity obtained by recasting a DTQW as a relativistic-like equation is proportional to $\sqrt{\cos\theta}$. 

Figure \ref{fig:4} presents MID and QD as a function of time for two different
noise levels. They present a similar degree of sensitivity (with fluctuations roughly in tune with
magnitude) and an expected overall reduction with noise. However, 
MID shows a {\it rise} in the regime $t \sim 10$ to $t\sim 60$, 
for the noise parameter $\lambda=0.01$, which would
clearly be unphysical for an indicator of non-classicality, as corroborated by
the monotonic fall of QD in this regime. 
This pathological behavior can be attributed to the non-optimization 
over local measurements in MID. It may be predicted that if the
optimization were performed, the resulting ameliorated MID
\cite{GPA10} would show monotonically decreasing behavior.
If one could analytically isolate the class of states for which MID applied to a DTQW shows such pathological behavior, and we are able to
confirm that the specific instance of walk dynamics 
does not involve such states, then
presumably one could still employ MID as a useful and easy-to-compute
indicator of quantumness \cite{SBC10}.

\section{Conclusion}
\label{conc}

Noisy quantum walks have been studied from
the perspective of comparing MID and QD as indicators of non-classicality,
when applied to linear and cyclic DTQWs. MID acts as a loose
upper bound to QD, sometimes properly reflecting even fine trends  
in the latter's behavior. However, there are regimes where
it obviously manifests artifacts due to the lack of optimization over
local measurements. 
\acknowledgments 
The numerical simulations for this work were done using the free {\tt numpy} numerical package for the Python language.


\end{document}